Astrophysical Explosions: From Solar Flares to Cosmic Gamma-ray Bursts

J. Craig Wheeler
Department of Astronomy, University of Texas at Austin

Abstract

Astrophysical explosions result from the release of magnetic, gravitational, or thermonuclear energy on dynamical timescales, typically the sound-crossing time for the system. These explosions include solar and stellar flares, eruptive phenomena in accretion disks, thermonuclear combustion on the surfaces of white dwarfs and neutron stars, violent magnetic reconnection in neutron stars, thermonuclear and gravitational collapse supernovae and cosmic gamma-ray bursts, each representing a different type and amount of energy release. This paper summarizes the properties of these explosions and describes new research on thermonuclear explosions and explosions in extended circumstellar media. Parallels are drawn between studies of terrestrial and astrophysical explosions, especially the physics of the transition from deflagration to detonation.

Keywords: neutron stars, black holes, supernovae, gamma-ray bursts, deflagration, detonation.

I. Introduction

Astrophysics is an observational science; astronomers carefully observe distant objects and events, but they cannot bring them into the lab to perform controlled experiments. For this reason, it is all the more important to make connections between terrestrial physics and the physics that is appropriate to astrophysical contexts. Study of explosion dynamics in both astrophysical and terrestrial contexts reveals common physics; study of the physics in one context can complement studies in the other. The release of energy in a magnetic pinch has much in common with solar flares. The propagation of shocks induced by high energy density lasers mimics behavior expected in astrophysical shocks. Thermonuclear bombs have their parallels in thermonuclear explosions of white dwarf stars. Some stars are expected to drive a "tamper" or "pusher" shell of matter inward to rapidly compress matter and trigger an explosion, in a manner reflected in the Teller-Ulam design of a thermonuclear device such as the W88 warhead or as invoked in the principle of laser implosion fusion. A primary example of astrophysical/terrestrial connections, that we will discuss below, is the topic of detonations and, in particular, the deflagration-to-detonation transition (DDT). Experimental and theoretical investigations of the propagation of terrestrial detonations has informed the study of astrophysical detonations, and examining the possibility of unconfined DDT in astrophysical combustion has led to the search for general mechanisms of DDT. DDT is thought to be involved in the thermonuclear explosions of stars that produce Type Ia supernovae. This type of supernova was used to discover the apparent acceleration of the Universe, one of the most profound issues facing modern astronomy and cosmology.

In astrophysical contexts, an "explosion" implies the release of energy on a timescale, $E/\partial E \partial t$, that is comparable to or shorter than the dynamical timescale, $R/v$, of the system, where E is a characteristic energy density, $\partial E \partial t$ its rate of change due to energy input, R is a characteristic size of the system and v is a characteristic velocity. The criterion $E/\partial E \partial t \leq R/v$ means that the system does not have time to mechanically respond quasi-adiabatically to the input of energy. The dynamical timescale is often the sound-crossing time for the system, $R/c_s$. In other circumstances the gravitational free-fall time, $R/v_{ff}$, is appropriate, but the sound-crossing time and the free-fall time are comparable for gravitationally-bound systems. This definition of explosion qualitatively pertains to both terrestrial and astrophysical explosions, although the word "explosion" is not normally used in technical combustion literature.

This paper will review a large range of astrophysical phenomena at a level suitable to introduce these phenomena to experts in terrestrial combustion. No attempt is made to provide deep mathematical or physical background, but relevant references are provided for readers interested in delving more deeply. The topics discussed cover a broad range of fields within astrophysics with each being the subject of a large variety of sub-communities. The author has contributed directly to work on most of the topics discussed here and pondered the remainder. Some of these topics are presented at a popular level in Wheeler (2007). The work outlined here covers decades of effort, but current and new work by the author and his collaborators will be discussed where appropriate.

The panoply of phenomena discussed here is associated with a huge range of spatial and temporal scales, from femtometers to gigaparsecs and from microseconds (or less!) to billions of years. Relevant densities range from nuclear densities of $10^{14}$ g cm$^{-3}$ to the rarified gas between stars with scarcely one particle per cubic centimeter or $10^{-24}$ g cm$^{-3}$. Atomic, nuclear, and plasma physics are all intimately involved. Turbulence and transport -- of particles, photons, and neutrinos -- are critical ingredients in virtually all the problems and on all scales. Astrophysical explosions tend to fall into three general categories according to the critical energy-releasing ingredient: magnetic fields, thermonuclear reactions, or gravitational collapse.

The Big Bang is commonly referred to as an explosion, although the Big Bang was of such a special nature that it does not fit the definition given in the Introduction. The Big Bang essentially represented the beginning of space and time and hence the beginning of physics as we know it. How that came about and the characteristics of any state that preceded the epoch of the Big Bang are yet beyond the reach of physics, if not enlightened speculation. From that first state, an incredible "creation story" emerges. The Big Bang represents a sudden investment of energy on the scale of the Planck time, about $10^{-43}$ seconds. Initially, space and the matter and energy that inhabited it, were subject to infinitesimal quantum perturbations. The distribution function of these perturbations controls the subsequent evolution of material in the Universe. These perturbations were Gaussian to the best of our knowledge, but this distribution is under intense scrutiny. There was a short interval of exponentially large expansion, the period of "inflation," and then the return to a more sedate expansion. The initial quantum perturbations left their imprint on what is termed Dark Matter. Dark Matter is thought to be particles that interact by the weak nuclear force and gravity, but that are not electrically charged and that do not interact by the strong nuclear force. The Dark Matter comprises about 23% of the mass/energy content of the Universe, whereas Baryonic matter – "ordinary" matter consisting of electrons, protons, and neutrons - is only about 5%. Thus Dark Matter dominates the evolution of gravitating matter in the expanding Universe. According to theory and simulations, the Dark Matter gravitationally agglomerated into concentrations and voids. Baryonic matter, a minor, but critical, constituent of the observable Universe, fell into these concentrations of gravitational attraction, compacted, and formed all the luminous structure we see today: planets, stars, galaxies, and clusters of galaxies. After the first instant of the Big Bang, the behaviors of expanding space and the Dark and Baryonic matter within it can be calculated and, though there are many issues to be resolved, the results adhere remarkably well to what we observe. This is the background in which we can discuss more classical astrophysical explosions.

## II. Solar Flares and Flare Stars

The nearest astrophysical explosions are solar flares that occur on the surface of the Sun (Tandberg-Hanssen & Emslie 2009; Figure 1). Solar flares have a direct impact on space weather and implications for the human space flight program. Solar flares tend to have a log normal distribution of energy, but the typical energy of large flares is about $10^{33}$ ergs, which is a substantial fraction of the luminosity of the quiescent Sun in one second. Solar flares are the sudden deposition of magnetic energy into the heat and kinetic energy of associated plasma. The physics of this sudden transition is not well known, but presumably involves instabilities, catastrophic changes of state that lead to the reduction of the energy of the magnetic state (Kagan & Mahajan 2010). The principal mechanism is thought to be magnetic reconnection. The magnetic field is fed from within the Sun by dynamo processes (Priest 1984). These are most popularly thought to have their origin in the solar tachocline, the shearing boundary between the outer convective and inner radiative portions of the Sun. The origin of the magnetic field and its role in causing solar flares is obfuscated by our lack of understanding of the rotational state of the Sun. Basic models predict that stars should rotate on cylinders and that convective regions that can rapidly exchange angular momentum should be in solid body rotation; yet the convective envelope of the Sun rotates on cones, as revealed by solar seismology (Howe 2009; Figure 2).

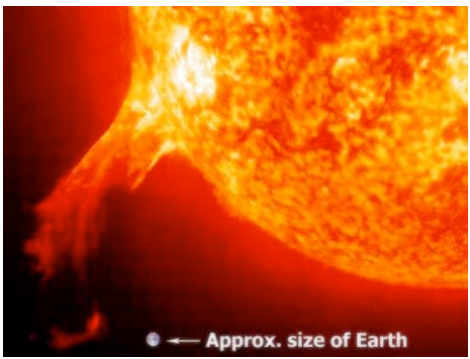

Figure 1 - The largest solar flare in four years occurred on February 17, 2011. Solar flares are important ingredients in space weather and the associated plasma physics has applications in controlled fusion. Courtesy NASA.

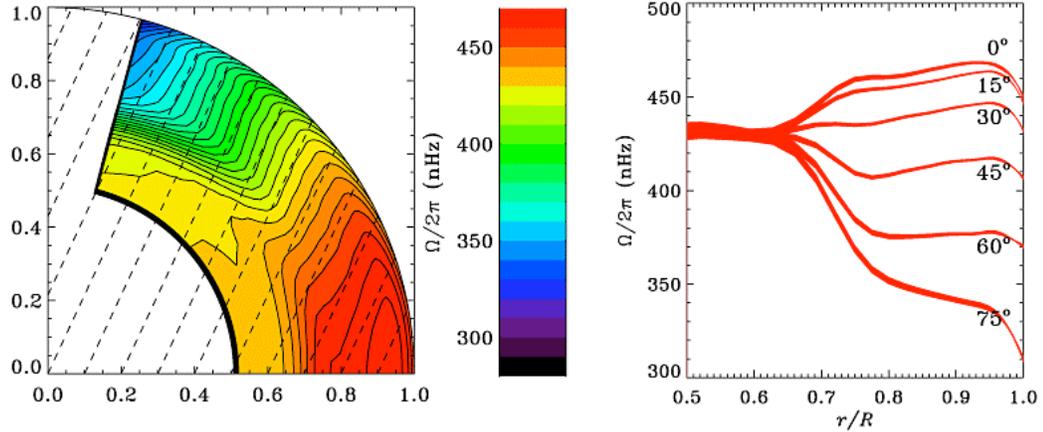

Figure 2 – Contours of constant angular velocity, Ω, in the outer portions of the Sun (left) and the profiles of Ω along lines of constant latitude (right). From Howe (2009).

Figure 2 shows the mean contours of constant angular velocity, Ω, in the outer portions of the Sun and the distribution of Ω along constant latitudes. The matter in the outer 5% of the solar radius and all the matter above about 50 degrees latitude have a negative gradient of the angular velocity. In the absence of stabilizing buoyancy effects in the convective envelope (beyond a fractional radius of about 0.6), a negative gradient in Ω is the criterion for the magneto-rotational instability (Velikov 1959; Chandrasekhar, 1960; Balbus & Hawley 1989). We are investigating the possibility that the solar convection zone itself can create some magnetic flux by means of the magneto-rotational instability (Kagan & Wheeler 2011).

We observe the analog of solar flares on other stars (Gurzadyan 1984; Hartman et al. 2011). In some circumstances, the results are spectacularly larger than for the Sun. Energies up to $10^{39}$ ergs have been estimated. These very large flares tend to occur on stars with less mass than the Sun where the convective envelope occupies an even larger proportion of the stellar mass. Stars like the Sun have been seen to produce these large flares on rare occasions (Rubenstein & Schaefer 2000). It is not known whether the Sun has, can, or will do so.

### III. White Dwarfs and Neutron Stars in Binary Star Systems

A. Cataclysmic Variables

Cataclysmic variables are white dwarf stars orbiting a companion star that transfers mass to the white dwarf (Warner 2003; Frank, King & Raines, 2002; Figure 3). Because of the orbital motion and conservation of angular momentum, the matter lost from the companion star forms a flat accretion disk around the white dwarf. The matter in the disk is subject to a magnetic turbulent viscosity (probably related to the magneto-rotational instability; Balbus & Hawley 1989) and slowly spirals inward. Angular momentum migrates outward and is eventually lost from the disk through interaction with the companion or loss of a small portion of the mass. There are two qualitatively different mechanisms that lead to outbursts of radiant energy in cataclysmic variables, with attendant loss of mass in some circumstances. One mechanism involves the accretion disk and produces what is known as a dwarf nova. The other mechanism involves thermonuclear burning on the surface of the white dwarf itself, resulting in phenomena known as classical novae and recurrent novae.

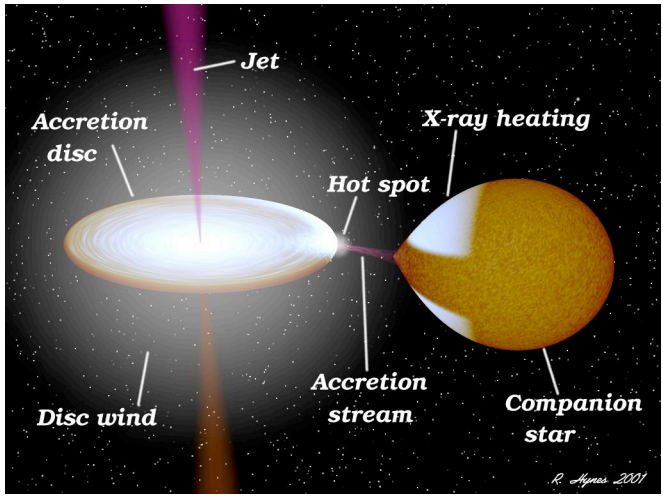

Figure 3 - Components of a cataclysmic variable: mass-transferring companion star, hot spot where transfer stream collides with the accretion disk, accretion disk and white dwarf star. The white dwarf is the size of the Earth and too small to be seen on the scale of this reproduction, but is in the center of the disk at the base of the jet. The hot accretion disk can heat the front surface of the companion star, and also eject winds and jets. Courtesy Rob Hynes: http://www.phys.lsu.edu/~rih/

The outburst in the accretion disk leading to a dwarf nova is not really an explosion, *per se*, in the sense that energy is released on a viscous timescale, not a dynamical, sound-crossing, timescale in the disk, but the result is dramatic (Meyer & Meyer-Hofmeister, 1981; Cannizzo, Ghosh & Wheeler 1982; Faulkner, Lin & Papaloizou, 1983). In certain ranges of the rate of mass transfer through the disk, the disk will start in a cold, neutral, semi-transparent state with optical depth less than or of order unity. Any heat energy generated by viscous processes is radiated away and the viscous transport of mass is slow. In this state, the disk cannot transport matter through the disk and onto the white dwarf at the rate mass is added to the disk. The disk thus accumulates mass in its outer regions. Eventually, as the density increases, the disk becomes opaque (optical depth greater than unity), traps thermal energy that is released by the viscous processes, and heats up. As the disk becomes ionized the opacity rises, the disk traps heat even more efficiently, and the viscosity, which increases with temperature, rises. The result is a thermal instability in the disk that triggers a heating wave that results in a redistribution of the disk mass and a sudden release of gravitational energy. The result, a dwarf nova, is characterized by a flare of luminosity by perhaps a factor of 10 – 100 that lasts for several days. Although there is wide variation, the characteristic energy liberated is of order $10^{40}$ ergs. Little mass is lost, but the redistribution of the disk mass and the deposition of some of the mass onto the white dwarf result in a thinning of the outer parts of the disk, a return to semi-transparency, and a return to the initial cold, neutral state. In this state, the disk becomes dim and begins to store matter to repeat the process on a timescale of a few weeks.

The class of cataclysmic variables that involves explosions on the surface of the white dwarf lead to classical novae or recurrent novae, depending on the physics and the associated repetition timescale (Warner 2003; Frank, King & Raines, 2002). For classical novae, hydrogen-rich matter is deposited on the surface of the white dwarf on a modest timescale that allows the added mass to cool and form an electron-degenerate layer. In such a layer, the pressure is insensitive to the temperature. The accumulating layer of hydrogen will eventually reach conditions of density and temperature at which the hydrogen begins thermonuclear burning. Because the pressure is insensitive to temperature, the material begins to heat, but there is no dynamical response and therefore a muted tendency to expand and cool adiabatically. Because the hotter matter burns more quickly, ignition under conditions of electron degeneracy leads to a thermal instability. By the time the hydrogen layer is hot enough that the degeneracy is lifted, the thermonuclear burning timescale is shorter than the dynamical timescale on the surface of the white dwarf and an explosion ensues. This explosion typically releases ~ $10^{46}$ ergs of energy and produces an optical outburst that makes the system ~$10^5$ times brighter than it is in its quiescent state. The hydrogen layer is about $10^{-4}$ times the mass of the Sun. This mass is blown into space, along with some of the matter from the white dwarf, in an outburst lasting about a month. Models predict that such explosions recur on a timescale of about 100,000 years (Starrfield, Truran & Sparks 2000).

Recurrent novae are a closely related phenomena, but of a milder sort. Models predict that they occur when the white dwarf has larger mass and hence larger surface gravity than classical novae (Starrfield, Sparks & Truran 1985). In this circumstance, the layer of hydrogen added to the surface of the white dwarf by mass transfer tends to retain the energy released by impact on the surface and the hydrogen layer tends to be sufficiently hot at a given density that it is not electron degenerate. When the layer of hydrogen ignites, this matter tends to expand and cool, but energy is still released on a rapid, thermal timescale. The release is less violent, and most of the matter is retained on the white dwarf. This matter eventually joins the inner, degenerate white dwarf material, effectively adding mass to the white dwarf. This may have other consequences, as will be described below.

B. Neutron Star and Black Hole Binary Transients

While astronomers give other names to phenomena associated with neutron stars and black holes in binary systems - X-ray novae, X-ray bursters, soft gamma-ray repeaters - such binary systems can have analogous phenomenological behavior to the cataclysmic variables described in the previous subsection (Tanaka & Lewin 1995; Chen, Schrader & Livio 1997). The instability associated with the accretion disk described above for dwarf novae is also thought to happen in the accretion disks around neutron stars and black holes in mass-transferring binary systems. Again, the phenomena are not true explosions in a dynamical sense, but are, nevertheless dramatic outbursts of energy. With the very high gravity near the surfaces of neutron stars and event horizons of black holes, the associated accretion disks get extremely hot during the thermal instability and emit energy in the X-ray band. These outbursts are characterized as X-ray novae and typically last for weeks, with a recurrence time ranging from years to decades. The energy released is of order $10^{45}$ ergs.

Since neutron stars have solid surfaces, material can collect on them and phenomena related to classical and recurrent novae can occur, although the higher gravity can alter the phenomenology. The class of explosions called X-ray bursts is thought to result from hydrogen-rich matter collecting and then rapidly burning on the surface of the neutron star (Grindlay, et al. 1976; Lewin, van Paradijs & Taam 1993). These outbursts rise to maximum luminosity of ~ $10^{38}$ erg s$^{-1}$ on timescales of seconds and decay on timescales of seconds to hundreds of seconds, emitting a total of ~ $10^{39}$ ergs. Models show accreted matter will undergo stable thermonuclear burning in the hydrogen, converting it to helium. In certain ranges of parameter space, however, the helium proves dense enough at temperatures where helium burns to become electron degenerate (Joss 1978). Thermonuclear ignition of helium is therefore unstable and leads to rapid burning on a dynamical timescale. Helium burning yields less energy per unit mass than hydrogen burning and the surface gravity of the neutron star is much greater than that of a white dwarf. The result is that the resulting helium explosion is substantially confined to the surface of the neutron star with a flare of X-ray emission, but rather little mass loss. There are suggestions that the helium burning may occur in shock-driven detonations that propagate around the periphery of the neutron star (Weinberg & Bildsten 2007).

Another phenomenon occurs near the boundary of the accretion disk and the magnetic field of the neutron star. Plasma from the disk apparently penetrates the field in blobs, falling inward and striking the surface of the neutron star. Since this process liberates gravitational energy on the dynamical timescale of infall, it might be considered a legitimate explosion. The observed phenomenon is associated with rapid bursts in X-rays from some binary neutron star systems (Lewin et al. 1976; Lewin, van Paradijs & Taam 1993)

Soft gamma-ray repeaters emitting energetic photons up to several MeV represent another example of violent explosions associated with neutron stars (Cline et al. 1980; Thompson & Duncan 1995). These sources rise on timescales of a fraction of a second to peak luminosities ~ $10^{45}$ erg s$^{-1}$ that radiate most of the energy The peak is followed by a decaying tail lasting several minutes that often reveals modulation at the rotational period of the neutron star. The total energy emitted ranges from ~ $10^{44}$ to $10^{46}$ ergs. In this case, models show that the outbursts result from dramatic rearrangement of the magnetic field configuration, presumably the result of field reconnection, in especially highly-magnetized neutron stars. Field energy is pumped into hot plasma trapped in the magnetic field that radiates in the spectral regime associated with low-energy, hence "soft," gamma rays. These bursts are thus distant and highly energetic cousins of solar flares. One of these bursts was so strong that it required the complete rearrangement of the magnetic field structure, e.g. something like the swapping of the north and south magnetic poles (Hurley et al. 2005).

While black holes can be the site of accretion disk instabilities, hence the term "black hole X-ray novae," they cannot display any of the other phenomena just ascribed to neutron stars that require the presence of a solid surface with perhaps an embedded magnetic field. The inner regions of the accretion disks around black holes do reveal "flickering" in their X-ray emission on millisecond time scales that is associated with blobs of plasma orbiting near the innermost dynamically stable portions of the disk (any closer and General Relativity predicts there are no stable circular orbits). These inner portions of the disk and plasma blobs may be associated with the magnetic structure of the disk and reconnection and so may also be distantly related to solar flares.

IV. Supernovae

Supernovae are explosions that represent the dynamical disruption of an entire star. They are typically $1 - 10 \times 10^9$ times brighter than the Sun and can briefly rival the brightness of their host galaxy with luminosities ranging from $\sim 10^{41}$ to $10^{44}$ erg s$^{-1}$. They typically produce $10^{49}$ to $10^{50}$ ergs of radiant energy and $\sim 10^{51}$ ergs of kinetic energy. Some produce 100 times more energy in neutrino output. As illustrated in Figure 4, supernovae typically rise to peak light in a couple of weeks and fade over the course of weeks or months, although some brighten and fade more slowly. There are two basic mechanisms: rapid thermonuclear burning and gravitational collapse. Important clues to the nature of the explosion come from the composition of the ejected material. Elements that are ejected in supernovae are frequently those of the "alpha chain," elements that have equal numbers of protons and neutrons, but differ from one another by four nucleons, effectively a helium nucleus or alpha particle. Common elements with these characteristics are carbon, oxygen, neon, magnesium, silicon, sulfur, calcium and titanium. The amount and distribution of these and other elements often give important clues to the nature of the explosion. The observed properties of supernova explosions and their compact and extended remnants can generally be placed in two categories corresponding to those mechanisms.

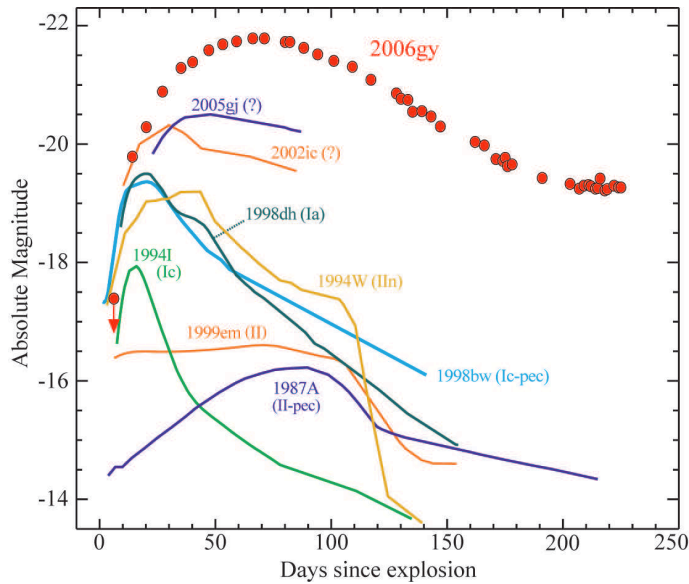

Figure 4 – illustration of the light curves of a variety of supernovae. The magnitude scale (historically related to the response of the human eye) is a logarithmic scale, with one magnitude being a factor of about 2.5 different in brightness, and smaller, more negative, numbers representing brighter events. From Smith et al. (2007).

A. Type Ia – Exploding White Dwarfs

One common type of supernova is called Type Ia (SN 1998dh in Figure 4). The remnant of the Type Ia witnessed by Tycho Brahe in 1572 is shown in Figure 5 and a more contemporary extragalactic example, SN 1994D, is shown in Figure 6. Type Ia are generally associated with "old" portions of host galaxies, implying that the progenitor star has lived a long time, typically billions of years, before exploding. The expanding ejecta of Type Ia show no spectroscopic evidence for either hydrogen or helium, the most common elements in the Universe. This means that the progenitor star must have consumed or expelled essentially all its hydrogen and helium before exploding. Spectra of Type Ia do show alpha-chain elements of intermediate mass - oxygen, magnesium, silicon, sulfur, calcium - in their outer layers near maximum light (Filippenko 1997; Wheeler & Benetti 2000). Several months after maximum the ejected material, presumed to represent the composition of the central portions of the ejecta, are rich in iron.

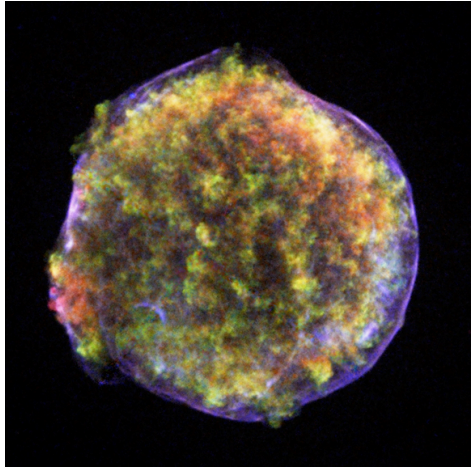 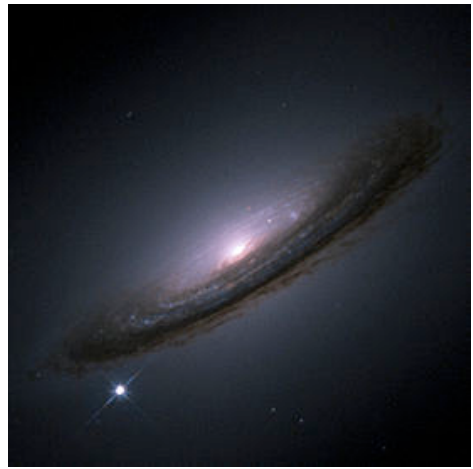

Figure 5 - False color X-ray image of the expanding remnant of the Type Ia supernova observed by Tycho Brahe in 1572. Courtesy of the Chandra X-ray Observatory.

Figure 6 – Type Ia supernova SN 1994D (the fourth supernova discovered in 1994), lower left, in the spiral galaxy NGC 4526. Courtesy High-Z Supernova Search Team. HST. NASA.

The most plausible progenitor star for Type Ia supernovae is an electron-degenerate white dwarf star (Hoyle & Fowler 1960; Hillebrandt & Niemeyer 2000). Models that best match the observations correspond to white dwarfs composed of carbon and oxygen. Models show that carbon/oxygen white dwarfs are produced by stars of moderate mass, those with less than about 8 times the mass of the Sun when they begin burning hydrogen. When the central hydrogen burns out, a core of helium is formed. In this moderate mass range, thermonuclear burning of helium will yield an electron-degenerate core composed of roughly equal parts of carbon and oxygen (Paczynski 1971).

To trigger the explosion, the white dwarf needs to acquire a mass near the maximum that can be supported by degenerate electrons, the Chandrasekhar mass of ~ 1.4 solar masses (Chandrasekhar 1931). At that point, carbon can ignite thermonuclear reactions. Because of the great electron degeneracy associated with these conditions, the initial release of energy heats the matter and strongly increases the rate of thermonuclear burning, but does not raise the pressure. By the time matter is sufficiently hot that the electrons are no longer degenerate and the pressure increases with increasing temperature, the burning timescale is comparable to the dynamical timescale. Models predict that the white dwarf is completely disrupted and observations of certain young supernovae in the Milky Way Galaxy are consistent with that prediction (Figure 5).

There are two plausible ways in which a white dwarf can be brought to the Chandrasekar mass to trigger this process. In one, the white dwarf orbits a companion star that is still burning hydrogen in its center or has not advanced much beyond that phase. In that circumstance, the companion transfers mass rather slowly, and the white dwarf grows slowly in mass. The combination of the long lifetime of the binary system before the mass transfer begins and moderate mass transfer rates can, in principle, give a long delay time before the explosion, hundreds of millions or billions of years. The other widely-considered model is one in which two white dwarfs are created in a binary system. In this case, the two stars emit gravitational radiation generated by the orbital motion and will spiral together. This process alone can take billions of years. When the white dwarfs get within the gravitational influence of one another, one can be rapidly disrupted and its mass deposited on the other white dwarf. This process can bring the remaining white dwarf to the Chandrasekhar mass (or higher) and trigger carbon ignition and explosion. Arguments pro and con are directed at both of these evolutionary mechanisms, and there is no general agreement as to which is correct. Both mechanisms may play some role. This author thinks that the preponderance of the data, especially that of the evolution of the spectrum over the course of the UV, optical and infrared display, is most consistent with the model based on the first scenario.

For the model where the white dwarf grows slowly to the Chandrasekhar mass, there are a combination of theoretical, simulation, and observational arguments that point to the detailed nature of the thermonuclear runaway. As the white dwarf grows in mass, the thermonuclear burning of carbon first proceeds slowly, but at some point, the rate of production of energy becomes more rapid than energy can be carried off by various neutrino production processes. At first the burning portions are unstable to thermal convection and the heating produced by the carbon burning is temporarily controlled by the outward convective transport of energy (Arnett 1969). Eventually, as mass is added to the white dwarf and conditions become more extreme, the burning rate becomes comparable to the dynamical timescale. At this point, the rapid release of energy triggers a turbulent deflagration driven by the Rayleigh-Tayler instability. This turbulent burning proceeds in a wrinkled flame regime with a net speed considerably in excess of the laminar flame speed. Pressure waves from the burning proceed in advance of the burning front and lead to the acceleration and expansion of the outer layers of the white dwarf. These outer layers expand in such a way that they become sonically decoupled from the burning. The turbulent deflagration can produce a substantial amount of energy, enough to unbind the white dwarf, but models show that the outer portions never burn; they are ejected as unburned carbon and oxygen in a rather meager explosion (Gamezo et al 2003). There are several objections to this model. The modest explosion does not quite give the observed velocities. The turbulent burning mixes some unburned carbon to deep layers with small velocity and carbon is not observed with the predicted small velocities in the ejecta. Finally, the outermost portions are predicted to be composed of unburned carbon and oxygen. Careful spectroscopic search for the carbon in the ejecta, in particular, has shown that most Type Ia supernovae do not have the predicted unburned carbon in the outer layers (Marion et al. 2006).

The models that seem to work best are ones in which there is a spontaneous deflagration to detonation transition (DDT; Gamezo, Khokhlov & Oran 2005). In current models, the condition for this transition is a parameter in the model, but with an appropriate choice of the density at which the DDT occurs, the resulting models bear remarkable resemblance to the observations (Höflich et al. 2006). The detonation catches up with and burns the outer layers, eliminating the outer unburned carbon that plagues the pure deflagration models. Numerical simulations also show that the detonation can penetrate inward and remove the carbon stirred inward by the turbulence. Since in this class of model the whole star is consumed by thermonuclear burning more energy is released, and the predicted velocity profile of the ejecta closely matches the observations. The detonated matter is predicted to be composed of the alpha chain elements oxygen (that is itself produced by burning carbon), magnesium, silicon, sulfur, and calcium, just as the observations require (Höflich et al. 2006).

The model based on the disruption of one white dwarf in a pair has not been subjected to the same level of quantitative radiative transfer analysis. It remains unclear whether or not such a model can reproduce the observed spectral evolution as well as the DDT model just described, but recent attempts are less than satisfactory (Fryer et al. 2010). Both models predict that the innermost ejecta that have undergone the most thorough nuclear processing will be composed of iron and iron-like elements. There are other observational constraints that have not been adequately addressed by either model. The outer layers of the ejecta of Type Ia supernovae are observed to be polarized and hence asymmetric in some way (Wang & Wheeler 2008). There is also a shell of matter moving at about 50% higher velocity than the photosphere that seems to represent the collision of the outer layer of the white dwarf with some external layer of matter. Both classes of models have the potential to explain these effects, but neither has done so quantitatively.

A major goal of current research in Type Ia supernovae is to confirm that the outer layers have been subject to a detonation. A new window into this research has been opened by the capabilities to discover the supernovae soon after the explosion so they can be tracked on their rise to maximum (Figure 4). This work has led to the discovery that a large majority if not all Type Ia show evidence for a high-velocity, optically-thin, shell moving at 20,000 – 30,000 km s$^{-1}$. These velocities are considerably higher than for the matter at the photosphere where the expanding ejecta become optically thick as viewed from the outside and have typical velocities of 12,000 km s$^{-1}$. The expanding ejecta quickly (on a timescale of minutes) settle into approximately homologous expansion with $v \propto r$ so that the higher velocity matter must also be at considerably larger radii than the matter at the photosphere. This high-velocity shell of matter shows emission from calcium that may be present in the gas from which the stars were born, but recent work has shown that it also contains silicon that must be produced and ejected in the explosion (Marion et al. 2011; Figure 7). A possible explanation for these high-velocity features is the impact of the detonated ejecta on a shell of circumstellar matter. Beside carrying information on the nature of the explosion, this shell of matter may also bring long-sought information on the nature of the binary star system that leads to the explosion. Another possibility is that the high-velocity features are formed as burned material accelerates unburned material in the white dwarf, with circumstellellar matter playing a small role. As for terrestrial detonations (see the contribution by Kessler, Gamezo &

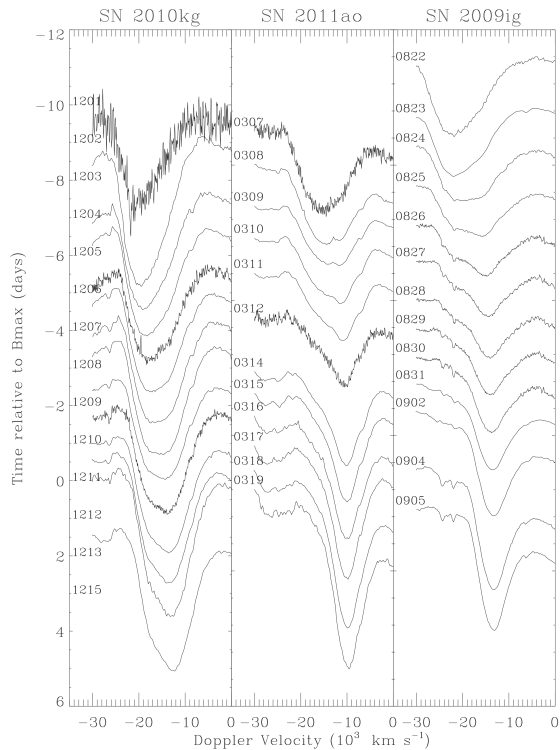

Figure 7 - A spectral absorption feature due to singly ionized Silicon as a function of velocity and time for three recent Type Ia supernovae. Numbers by each spectrum indicate the date it was acquired. These sequences all start about a week before maximum light (see Figure 4) and continue to somewhat after maximum. Note the shift in the velocity corresponding to the absorption minimum and the suggestion, confirmed by quantitative analysis, of two components, one that is associated with v ~ 20,000 km s$^{-1}$ that fades and one at v ~ 12,000 km s$^{-1}$ that persists. The high-velocity component holds clues to the burning by detonation of the outer layers. From Marion et al. 2011, unpublished.

Oran in this volume), astrophysical detonations produce a characteristic cell structure as they propagate. The nature of the way the burning dies in the low density outer layers of the white dwarf depends on the detonation cell structure associated with the various fuels. Carbon-detonation cell structure will dominate the outermost layers (Domínguez & Khokhlov 2011).

An important aspect of the initial deflagration of a white dwarf is that the burning of the thermonuclear fuels, carbon and oxygen, in the conditions predicted by models produces a substantial amount of $^{56}$Ni (Colgate & McKee 1969). This element has nearly the largest binding energy per nucleon of any element, but has an equal number of protons and neutrons. It is thus selectively produced in the rapid thermonuclear burning of fuels such as $^{12}$C and $^{16}$O, which themselves have equal numbers of protons and neutrons. The nickel is unstable and decays with a half-life of 6.6 days to produce $^{56}$Co that is also unstable and itself decays with a half-life of 77 days to produce $^{56}$Fe that is stable. The explosion of a white dwarf alone would not produce a luminous display because severe adiabatic expansion and cooling convert all the initial thermal energy of the explosion into kinetic energy of expansion. The subsequent deposition of energy from the decay of nickel and cobalt provide a crucial means to re-heat the ejecta to provide the characteristic optical display of a Type Ia supernovae.

Because of their nearly, although not exactly, uniform behavior, Type Ia supernovae proved to be powerful tools to determine distances to their host galaxies. The distance to a supernova is determined by comparing its known intrinsic luminosity to its observed luminosity using the complete cosmological expression from General

Relativity, but basically by invoking the inverse $r^2$ law for observed flux. The supernova distance determined in this way is independent of the recession velocity as determined by the redshift of the host galaxy. These two independent pieces of information, applied statistically to a large sample of Type Ia supernovae, revealed the remarkable, and deeply challenging, result that the Universe is accelerating, most often interpreted as the effect of a "Dark Energy" (Riess et al. 1998; Perlmutter et al. 1999). The supernova results coupled with other techniques (fluctuations in the cosmic microwave background radiation, the spatial distributions of galaxies, and the growth of clusters of galaxies) suggest that this effect is consistent with Einstein's Cosmological Constant in a Universe that is geometrically flat. That the Cosmological Constant should have the particular small, but non-zero, value it seems to have is a profound challenge to physics.

B. Core-collapse Supernovae

Many supernovae of various observational properties are thought to arise after a process of gravitational collapse (Fowler & Hoyle 1964; Colgate & White 1966; Bethe 1990). Stars with initial masses greater than 8 solar masses are predicted to form heavy-element cores that, for various reasons outlined below, evolve to a condition where the internal pressure cannot withstand the pull of gravity. Technically, this comes about when the increase in pressure due to an infinitesimal compression is not sufficient to maintain hydrostatic equilibrium in the compressed state). This condition corresponds to dynamical instability and the result is that the core collapses on a free-fall timescale, typically seconds. Figure 8 gives an X-ray image of the remnant of the "guest star" of 1054, the Crab Nebula, and Figure 9 gives a contemporary example of an extragalactic supernova, SN 1987A, known from its neutrino emission to have arisen by core collapse. The liberation of gravitational potential energy in the process of collapse to form a neutron star is predicted to generate about 100 times more energy than needed to produce observed supernova kinetic energies. The problem of several decades standing is that most of this potential energy is converted to the energy of neutrinos that escape the collapsing configuration nearly un-impeded. This has proven a very difficult problem to treat quantitatively.

Stars in the mass range between about 8 to 15 times the mass of the Sun may form electron-degenerate cores of oxygen, neon, and magnesium that are the products of carbon burning. If the outer envelope of the star is ejected in some manner (winds, violent oscillations, or by transfer to a binary companion), the core can be exposed as a white dwarf. If sufficient mass is added to such a stellar core still enshrouded in its host star or to a white dwarf counterpart by binary star mass transfer, the neon and magnesium undergo electron capture that reduces the electron degeneracy pressure. The loss of pressure support leads to dynamical collapse (Miyaji et al. 1980). Although such a configuration is composed of potentially combustible matter, the oxygen, neon, and magnesium, models predict that burning during the collapse is not sufficiently energetic to slow the infall substantially. The result is expected to be the formation of a neutron star. Because stars in the mass range of 8 to 15 solar masses have relatively little mass immediately beyond the collapsing core they explode relatively easily. Some models predict that neutrino deposition can lead to a modest explosion (Kitaura et al. 2006). There are statistical reasons to think that many supernovae come from this mass range, but there is no definitive evidence that the explosions specifically involve collapse of oxygen/neon/magnesium structures. The observable results would be different depending on whether the collapsing core were still surrounded by an extended envelope of matter or were a "bare" O/Ne/Mg white dwarf directly exposed to space.

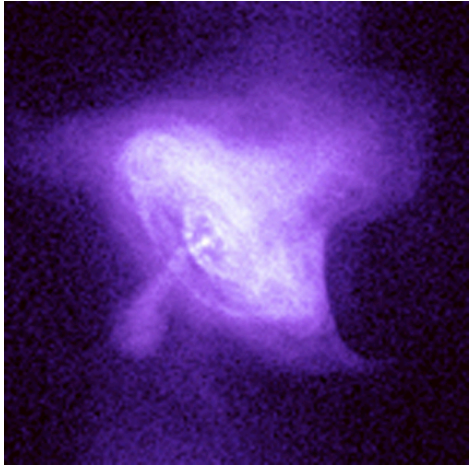

Figure 8 - False color X-ray image of the Crab Nebula, the remnant of the "guest star" of 1054. The point source in the center is the neutron star, a magnetic rotating pulsar. The surrounding nebula is the emission of synchrotron radiation by particles emitted from the neutron star. Courtesy of the Chandra X-ray Observatory.

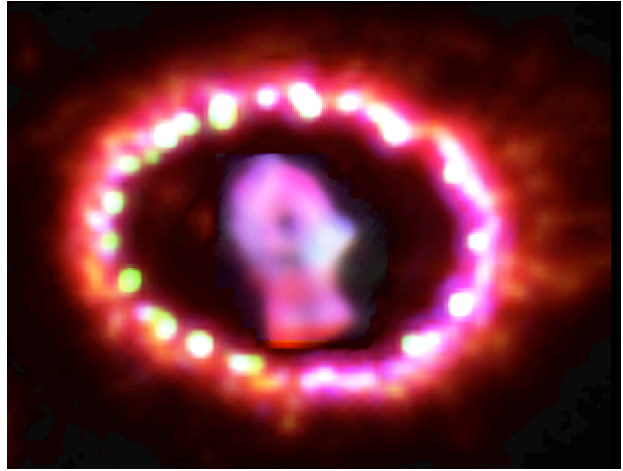

Figure 9 – Image of the remnant of SN 1987A in March 2011 (central elongated image) surrounded by the ring of matter expelled by the progenitor star prior to the explosion. The ring has now been struck by supernova ejecta producing the bright "pearl necklace" around its perimeter. Coutesy P. Challis, the SAINTS team and NASA.

Most of the theoretical work on core collapse has been in the mass range above about 15 solar masses. In this range, stars are predicted to form a core composed primarily of $^{56}$Fe that represents the maximum nuclear binding energy per nucleon. Further compression of this iron core triggers endothermic nuclear reactions that "soften" the equation of state so that the pressure does not increase sufficiently with density upon compression. This again leads to dynamical instability. The destabilized iron core collapses in free fall and nearly homogolously ($v \propto r$) to form a neutron star, or perhaps in some circumstances, a black hole, in about one second. At the surface of the neutron star, a shock wave forms that propagates outward into the infalling matter. The strength of this shock is tamped by dissociating the infalling iron and by neutrino losses (Bethe 1990). While this prompt shock must form, one of the few unambiguous conclusions in this complex area is that, at least in spherical models, this process cannot lead to explosion. Rather, the shock stalls within a few milliseconds and forms a standing shock at a distance of order 10 times the radius of the proto-neutron star. Spherically symmetric models fail to explode, but contemporary multidimensional simulations suggest that in some circumstances the flux of neutrinos from the neutron star may heat the regions just within the standing shock. This extra heating can re-accelerate the shock, driving it outward to cause the explosion (Nordhaus, et al. 2010).

One especially interesting recent development was the discovery that the standing accretion shock is intrinsically unstable to a standing accretion shock instability (SASI) that will drive strong deviations from spherical symmetry in the vicinity of the standing shock (Blondin & Mezzacappa 2007). This instability guarantees that the conditions near the shock are strongly non-spherical. Whether or not this instability aids the explosion by altering the dynamics or the neutrino deposition or by generating acoustic or magnetic flux, is under active investigation

Core-collapse supernovae have a variety of light curve and spectroscopic behavior. Figure 4 gives that light curves of SN1987A, SN1994I, SN1994W, SN1998bw, and SN1999em, all of which are thought to result from core collapse. The most common variety of core-collapse supernovae are called Type IIP. These are characterized by abundant evidence for hydrogen in the spectra and a light curve that is characterized by a "plateau" (hence the "P") lasting for about a month (SN1999em in Figure 4). These events represent an explosion within the extended envelope of a red giant star. Events called Type IIb represent similar explosions where much of the outer hydrogen envelope has been lost. Type Ib supernovae are even more hydrogen deficient, but still show evidence of underlying layers of helium. Type Ic show little or no evidence for either hydrogen or helium (SN1994I and SN1998bw in Figure 4). The latter three categories are thought to arise from massive stars that have lost part or all of their outer layers to "winds" from the surface of the star or to transfer of mass to a companion star in a binary system.

The principle evidence directly connecting the theory and observations of core collapse came from Supernova 1987A (Figures 4 & 9). The progenitor star had about 20 that we observed to be a star of about 20 solar masses prior to explosion. We directly detected the neutrinos associated with the core collapse and explosion (Arnett et al. 1989; McCray 1993). We clearly observe neutron stars and black holes within our own Milky Way galaxy and in a few nearby galaxies that are the compact remnants of the deaths of massive stars. In addition, we observe the old, extended remnants of massive star explosions within our Galaxy that were witnessed prior to the epoch of modern astronomy, but that have ejected large amounts of mass and left behind a compact remnant. Among these are the Crab Nebula (Figure 8) and the Cassiopeia A supernova remnants. For distant supernovae in other galaxies, the evidence is only indirect that a supernova has undergone core collapse to leave behind a neutron stars or black hole. The evidence that some supernovae arise in massive stars, and hence from theory must undergo core collapse, ranges from their explosion within regions of active star formation, a characteristic of massive stars, to direct measurement of the large mass of the ejecta.

Core-collapse supernovae commonly produce and eject $^{56}$Ni that is thought to come from the rapid, explosive, burning of silicon layers that surround the iron core. The bright plateau phase of Type IIP tends to mask this input, but it is seen in the "tail" of the light curve after the plateau phase. Type Ib and Type Ic arise in compact progenitors that, like Type Ia, deplete their original thermal energy in adiabatic expansion. As for Type Ia, the observed light display from Type Ib and Type Ic supernovae is thought to come entirely from the later deposition of radioactive decay energy. Some supernovae show very little evidence of radioactive decay: these are speculated to be associated with stars undergoing the collapse of O/Ne/Mg cores where there is no silicon layer to produce nickel or with those that collapse to form black holes, swallowing the radioactive species in the process.

Advancing technology, especially the ability of the Hubble Space Telescope, has enabled the detection of the progenitor stars of Type IIP supernovae by searching for evidence of the progenitor in CCD images obtained for other reasons prior to the eruption of the supernovae (Smartt 2009). This technique has so far been limited to Type IIP (also SN 1987A and one Type IIb) because the individual red giant progenitors can be detected, barely, in nearby galaxies. Evidence has accumulated that Type IIP supernovae come predominantly from stars with mass in the range about 8 to about 15 solar masses as expected from the demographics; progenitor stars, and hence their supernova progeny, are more abundant at lower masses. As remarked above, this implies that most SN IIP arise from electron-capture induced collapse, not iron-core collapse. The latter is not precluded, just more rare. The hydrogen-stripped progenitors of the Type Ib and Type Ic supernovae are expected to be relatively dim; they are hotter on their surface than the red giant progenitors of Type IIP, but have much smaller surface area. In any case, no progenitor of a Type Ib or Type Ic has been as yet detected.

An interesting aspect of the core collapse supernovae is that they are all significantly and routinely asymmetric (Wang & Wheeler 2008). There is a tendency for the ejecta to show a favored axis, but also evidence for departure from axial symmetry, including a segregation of different elements being blasted out in different directions. This behavior is associated with the underlying explosion mechanism. The asymmetries associated with the standing accretion shock instability might be manifest here or there may be asymmetries associated with the deposition of neutrinos. Another alternative is that the neutron star, which is surely rotating, may also be significantly magnetic, with magnetic pressure approaching the gas pressure (Wheeler et al. 2000; Akiyama et al. 2003). Rotating, magnetic configurations in other astrophysical contexts frequently generate magnetically driven jets. Another possibility is then that the observed asymmetries arise from the propagation of magnetic jets arising near the new neutron star and propagating out through the star.

C. Pair-instability Supernovae

Very massive stars, those exceeding 100 solar masses at birth, are predicted to get hot enough that the ambient photons in the interior, effectively gamma-rays, are sufficiently energetic to create electron/positron pairs (Rakavy & Shaviv 1967; Barkat, Rakavy & Sack 1967; Fraley 1968). The conversion of energy to rest mass rather than thermal energy also alters the equation of state. The pressure does not increase sufficiently with density upon compression to support the structure against gravity. Pair formation is yet another mechanism to "soften" the equation of state and destabilize the star. Models predict that this happens after stars have undergone their central helium burning and have formed massive cores composed primarily of oxygen. In models, the instability occurs in an off-center shell (Chatzopoulos, Dearborn & Wheeler 2011). The rapid contraction of this shell drives it inward like the tamper of the W88 warhead mentioned in the Introduction. This leads to the rapid compression and heating of the inner core of oxygen. Unlike the case of iron-core collapse, the oxygen in these stars is subject to strong energy release by rapid thermonuclear burning. The result is the prediction that the star is totally disrupted, leaving no remnant, but with the production of a very large mass of radioactive $^{56}$Ni, the decay of which could power the light output.

For decades, this model for pair-instability supernovae was regarded as "straightforward" as an explosion mechanism, but there was strong doubt that stars of sufficient mass actually formed in Nature. This perspective changed with the theoretical study of the first stars formed after the Big Bang when there were yet no heavy elements, beyond hydrogen and helium, in the ambient gas. Under these conditions, models predict that the stars that form will be especially massive so that the first stars might be especially susceptible to forming these pair-instability supernovae (Bromm & Larson 2004).

Another important development was the discovery in the distant, but still basically contemporary, Universe of a category of super-luminous supernovae that are relatively rare, but brighter by a factor of 10 to 100 than the other "normal" supernovae that have been described here (Smith et al. 2007; Quimby et al. 2007). Some of these super-luminous supernovae show evidence of high mass, but cannot be pair-instability supernovae. If their brightness derived from the radioactive decay of $^{56}$Ni, as demanded by the pair-instability model, they would require a greater mass of $^{56}$Ni than is allowed for the total mass of the ejecta. The ejecta mass is constrained by the width of the light curve that is a measure of the diffusion time. SN2006gy in Figure 4 is an example of a supernova with a long, slow, very bright light curve, but quantitative analysis revealed that it cannot to be a pair-instability supernova (Chatzopoulos, Wheeler & Vinko, 2011). The great luminosity of super-luminous supernovae like SN2006gy probably derives from the collision of the ejecta with a shell of matter previously ejected by the progenitor star (Smith et al. 2007; Chatzopoulos, Wheeler & Vinko, 2011).

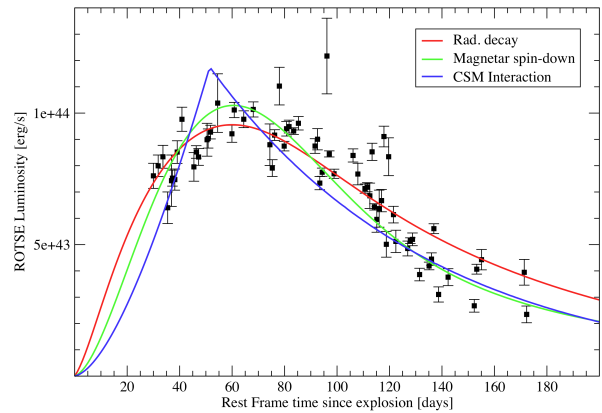

Figure 10 - ROTSE CCD-band light curve data for SN2010kd with three semi-analytical models for the light curve: one powered by radioactive decay as expected for a pair-instability model, one driven by the radiation of a spinning dipole as expected for a highly magnetized neutron star, a magnetar, and one where the power input derives from the collision of supernova ejecta with a circumstellar medium produced by the progenitor star before it exploded. The pair-instability model gives a plausible overall agreement with the data. From Chatzopoulos et al. 2011b, in preparation.

Several of these super-luminous events do, however, have all the characteristics expected of a pair-instability supernova. A recent example is SN2010kd, discovered by the ROTSE robotic telescope (Akerlof et al. 2003) at the University of Texas. Figure 10 shows the light curve of this supernova obtained by ROTSE and several models of the light curve (Chatzopoulos et al. 2011a; Chatzopoulos, Wheeler & Vinko, 2011; Chatzopoulos et al. 2011b). A pair-instability model powered by radioactive decay provides a plausible fit to the data.

V. Cosmic Gamma-ray Bursts

One of the most dramatic chapters in the study of astrophysical explosions has played out over the last decade (Woosley & Bloom 2006). Cosmic gamma-ray bursts were discovered in the 1960's by satellites launched to monitor nuclear test ban treaties. Unlike the soft-gamma ray repeaters described above, these sources never repeat. The bursts lasted for 10s of seconds, too short a time to be observed in other bands. For decades, the source of the bursts could only be located to within several degrees on the sky, an impossibly large volume to search once the flare had faded. The inability to obtain ancillary data meant that the distance and hence the intrinsic power was unknown. Free of this critical fetter, models ranged from relativistic dust grains colliding with the solar wind to vast explosions on the far side of the Universe. Evidence accumulated that the gamma-ray bursts were distributed isotropically on the sky, feeding the notion that they were distant cosmological events. In 1998, an Italian-Dutch satellite was instrumented with an X-ray detector that allowed a better localization of the event, within arc minutes on the sky. Quick work with a traditional ground-based telescope revealed that there was a lingering source of luminosity, an "afterglow" that enabled the precise location of the event to be determined (van Paradijs et al. 1997). That location proved to be a very distant galaxy, thus establishing that the gamma-ray bursts were cosmologically distant, billions of light years, and, indeed, intrinsically very powerful. The discovery of the afterglow phenomenon ushered in a new rush to discover these events. This involved nearly every observatory on Earth and both existing and newly-launched satellites.

A. Long Soft Bursts

Discoveries cascaded quickly. It was soon understood that to avoid unacceptably high optical depths for the gamma-rays that could produce their own cloud of electron/positron pairs, the associated flow of matter must be highly relativistic, with a Lorentz factor, $\Gamma > 100$ (Piran 2004). The study of gamma-ray bursts depends critically on the employment of Einstein's special relativity since both the intrinsic rest-frame time scales and associated length scales can be very different from those directly observed. In addition, the emitted power is subject to strong relativistic beaming in the direction of the flow.

Given the great distances, the intrinsic luminosity and integrated energy release of the gamma-ray bursts threatened to drive the subject into unprecedented physical regimes, comparable to the total annihilation of the rest mass of stars. This conundrum was also soon resolved with the recognition that the energy from the gamma-rays bursts is not emitted isotropically, but is strongly collimated (Rhoads 1997). This is a true collimation of the flow into a "jet" in addition to the Lorentz beaming due to the strongly relativistic motion. The jets are deduced to have typical opening angles of about 10 degrees. This collimation has two critical implications. One is that the energy demands are severely reduced compared to the assumption of isotropic emission. The energy involved turns out to be comparable to the energy of a typical supernova. For gamma-ray bursts, however, this energy is concentrated in a small amount of mass, roughly that of Jupiter, rather than into the mass of a star, as is characteristic of a supernova. The other important implication is that we do not see most of the gamma-ray bursts, those that are aimed by the collimation and relativistic beaming away from Earth; there are of order 100 bursts for every one that "points" at the Earth.

The next revealing development was that gamma-ray bursts were associated with regions of active star formation within their host galaxies. This suggested, indirectly, that they arose from the death of massive stars. The argument, also used for supernovae, is that massive stars live only a short time, typically millions of years. This does not give them time to drift away from their birth sites. The implication is that if a star dies near its birth site, a region of active star formation, it has likely arisen in a massive star. This logic set off a race to look for supernovae associated with gamma-ray bursts. The first hints were from an excess of light detected a about two weeks after the gamma-ray burst, the typical time needed for a supernova to reach peak light (Figure 4). Further study using spectroscopy revealed that the excess light was from a supernova of a very particular kind: Type Ic (Patat 2001; Stanek et al. 2003).

The supernovae associated with gamma-ray bursts were an especially interesting category of Type Ic. There was no evidence for hydrogen or helium, the operational definition of the class, but the residual features were strongly Doppler broadened compared to "normal" Type Ic. There is still debate as to whether this broadening is due to an excess of kinetic energy, to the effects of asymmetry such that we might be looking at the especially high-velocity component, or a mix of both effects. Not all Type Ic are broad lined; not all broad-line Type Ic are associated with gamma-ray bursts; but all supernovae associated with gamma-ray bursts are broad-line Type Ic. The reasoning is that the underlying mechanism, a relativistic jet, might occur more frequently in other types of supernovae, but that unless the outer envelope of the star has been stripped by winds or transfer to a binary companion, the jets cannot escape into space to produce the gamma-ray burst.

It is now understood that the afterglow is the result of the gamma-ray burst jet colliding with surrounding matter, either circumstellar matter shed by the progenitor or the interstellar matter of the host galaxy. This interaction plays out over weeks or months in the rest frame of the host galaxy, but can be reduced by the relativistic motions to days or weeks in the frame of the Earth. The nature of the gamma-ray burst itself remains under intense study. A popular model has been colliding shocks within the relativistic outflow (Figure 11). This picture has recently been challenged by new information from the Fermi satellite that detects gamma-rays from both the original burst and the afterglow (a technical challenge). These new data suggest that the old picture is incorrect (Kumar & Barniol-Duran 2010). There are also still major issues as to whether the flow of energy in the gamma-ray bursts is dominated by ordinary matter, baryons, or by magnetic fields and their associated reconnection. The new data seem to be pointing in the latter

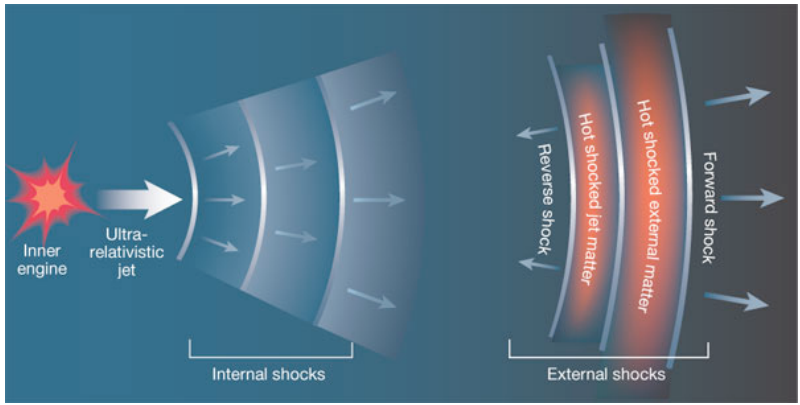

Figure 11 - Canonical model of the origin of a cosmic gamma-ray burst by internal shocks within a relativistic jet and the subsequent interaction of the jet with a circumstellar medium (courtesy T. Piran).

.
direction. Perhaps gamma-ray bursts are very, very large cousins of solar flares, where we began this discussion This leaves the remaining mystery of the fundamental source of the explosion. It occurs in the bare core of a massive star. Some "engine" deep inside powers both an explosion of the star and the production of a highly-collimated, relativistic jet of particles or perhaps magnetic fields. The combined need for high energy and highly relativistic motion has made the production of a black hole a leading contender. Models consider the formation of a rotating black hole and an associated accretion disk deep within the star. This combination could plausibly generate a baryonic or magnetic jet that could tunnel out of the star to produce the gamma-ray burst and afterglow (Woosley 1993; Kumar, Narayan & Johnson 2008). In this class of model, more traditional physics, neutrino deposition and associated effects, is invoked to cause the actual supernova explosion. A recently proposed alternative is that a standing accretion shock could form around the black hole that could, under proper circumstances, propagate out through the star triggering the explosion, with the jet again arising near the vicinity of the black hole (Lindner et al. 2010). An alternative model invokes the birth of a magnetar, the sort of highly magnetic neutron star associated with soft gamma-ray repeaters (and perhaps supernova asymmetries). With the magnetic field to tap and collimate the energy flow, this configuration might also both drive a supernova explosion and power a relativistic, perhaps magnetically-dominated jet (Bucciantini et al. 2008).

B. Short Hard Bursts

The study of gamma-ray bursts has revealed a second, but related phenomenon. The bursts that have just been described last for 10s of seconds and their bursts are characterized by relatively low-energy, "soft" gamma rays. These are now formally known as "long, soft bursts." Another category is characterized by shorter bursts, ranging from 0.1 to a few seconds. These bursts tend to have somewhat more energetic, harder gamma-ray spectra. They are thus known as "short, hard bursts." There is some overlap in this qualitative description, and the short hard bursts also do not repeat, but the short hard bursts clearly represent a separate physical category. Further study has shown that they also arise in distant galaxies, but the distances are less, hence the events are somewhat less energetic. Evidence for collimation is ambiguous. No short hard burst has ever been associated with a supernova explosion. While short hard bursts can occur in star-forming galaxies, they are not confined to that type of galaxy and do not concentrate in the star-forming regions themselves (Berger 2011). The conclusion is thus that they do not arise directly in the explosion of a massive star. They may arise from an originally massive progenitor, but they must live long enough before producing an explosion to drift far from their birth site, typically billions of years.

There are two main contenders to account for the short hard bursts. Perhaps the most popular is a model based on two neutron stars in binary orbit. Such a configuration would lead to an inspiral of the orbits due to production and loss of gravitational radiation. As for two white dwarfs, this is expected to lead to the disruption of the lower mass neutron star and the sudden dumping of its mass onto the more massive neutron star. Binary neutron stars are observed, the inspiral is inevitable given enough time, and the subsequent interaction will surely be a dramatic event (Ruffert & Janka 2001; Rosswog, Ramirez-Ruiz & Davies 2003). There remain, nevertheless, quantitative issues of whether or not this model will give the observed spatial distribution and whether or not the associated event will give the specific observed properties of a short hard burst. An alternative model is again one based on a magnetar. This model proposes that some O/Ne/Mg white dwarfs accrete enough mass to trigger the electron capture instability mentioned in Section IVB and then undergo collapse to form neutron stars. White dwarfs accreting in binary systems are likely to be rapidly rotating and that rotation may itself induce strong magnetic fields in the original white dwarf and in the neutron star. It is thus reasonable to consider that the collapse of a white dwarf could produce a rapidly rotating, highly magnetic neutron star. This neutron star might generate magnetic jets that could yield phenomena like the short hard bursts (Metzger, Quataert & Thompson 2008). Once again, this has not been proven quantitatively. The origin of the short hard bursts, as for the long soft bursts, remains a challenge to understand.

## VI. Conclusion

We witness astrophysical explosions that release magnetic, gravitational, or nuclear energy on a vast range of time, length, density and energy scales, from solar flares to cosmic gamma-ray bursts. Magnetic phenomena are probably much more ubiquitous in these explosions than has been widely appreciated, and in many cases all three energy sources are at play. It remains a challenge to incorporate all the physics on relevant scales of length and time to adequately simulate these events. For decades we were conceptually and computationally restricted to model many of these explosions as spherically symmetric. There is now a wide-spread understanding that all these phenomena are deeply and critically three-dimensional processes.

Numerical simulations will be critical to obtain deeper understanding. We will still depend on the art of choosing the right simulation. Despite Moore's law, we do not yet have the computational ability to fully address the problem of core collapse in all its collapsing, rotating, turbulent, magnetic, nuclear-reacting, neutrino-transporting glory. We need to understand the physics of magnetic dynamos, field reconnection and turbulence, the deflagration to detonation transition, and turbulent, reactive, radiative flows. Then we need to encapsulate that physics in appropriate simulations to address the vast array of observed astrophysical explosions.

JCW thanks the editors for the opportunity to construct this review. His research is supported in part by NASA, the NSF, DOE, and the State of Texas.